\documentclass{birkjour}
\usepackage{amsfonts}
\usepackage{latexsym}
\usepackage{amsmath,amssymb,graphicx}
\usepackage{tikz}
\usepackage[all]{xy}
\usetikzlibrary{arrows,shapes}
\usetikzlibrary{positioning}

\newcommand{\Kall}{\ensuremath K[D]}

\def\t#1#2#3{#1\stackrel{#2}{\xrightarrow{\hspace*{0.8cm}}}{#3}}
\def\tt#1#2#3{#1\stackrel{#2}{\xrightarrow{\hspace*{1.2cm}}}{#3}}

\newtheorem{theorem}{Theorem}[section]
 
 \newtheorem{lemma}[theorem]{Lemma}
 
 \theoremstyle{definition}
\newtheorem{definition}[theorem]{Definition}
 \theoremstyle{remark}

 \numberwithin{equation}{section}
\begin{document}

%
%
%
%
%
\title[First results on orbits of Darboux groupoid]{Orbits of Darboux groupoid \\
for hyperbolic operators of order three}
\author[Ekaterina Shemyakova]{Ekaterina Shemyakova}
\address{%
Department of Mathematics\\
State University of New York at New Paltz\\
1 Hawk Dr. New Paltz, NY 12561}
\email{shemyake@newpaltz.edu}
\subjclass{Primary 70H06; Secondary 34A26}
\keywords{Darboux groupoid, Darboux transformations, Laplace transformations, differential invariants, invertible Darboux transformations, Darboux transformations of type I}
\date{December 1, 2014}

\begin{abstract}
Darboux transformations are viewed as morphisms in a Darboux category.
Darboux transformations of type I which we defined previously, make an important subgroupoid consists of Darboux transformations of type I. 
We describe the orbits of this subgroupoid for hyperbolic operators of order three.

We consider the algebras of differential invariants for our operators. In particular, we show that the Darboux transformations of this class can be lifted to transformations of differential invariants (which we calculate explicitly).
\end{abstract}

\maketitle
\section{Introduction}

Darboux transformations are discrete symmetries of linear differential equations, ordinary or in partial derivatives,
and are also the basis of B\"acklund transformations of the non-linear theory. 
A description of all Darboux transformations for a given equation is an open problem. 
For the current state of the art see~\cite{ts:genLaplace05,tsarev06,Ganzha12,grinevich_novikov_2012discrete,ganzha2013intertwining,2013:factori:darboux:d,mikhailov2014darboux}.

The best studied Darboux transformations are those that are obtained from a number of linearly independent solutions
through the Wronskian formulas. In this way one can construct Darboux transformations for equations of a very general 
form~\cite{nimmo2010}. For practical purposes one would want to have other types of Darboux transformations too, since 1) the transformations of Wronskian type are not invertible;
2) the other popular special case of Darboux transformations - two Laplace transformations - is not in this class (in particular they are invertible).

Recall that the Laplace transformations are two transformations defined for a very special type of equations, namely, second order hyperbolic equations in two independent variables.
They are related to factorizations of the corresponding operator. The known Laplace transformation method is based on the fact that Laplace transformations can be described by transformations of the corresponding gauge invariants. 
After a long history~\cite{veselov:shabat:93,shabat1995,bagrov_samsonov_1995,adler1993,Veselov2003,Laplace_only_degenerate_2012,shem:darboux2,shem_2014_factorization_of_DT_for_factorizable}, 
the hypothesis that can be traced back to Darboux was proved~\cite{2013:factori:darboux:d} that the Darboux transformations of any order for such equations are either of Wronskian type,
or can be decomposed into a product of Laplace transformations. 

In~\cite{2013:invertible:darboux} it was first noted that 1) there are a number of examples of operators of third order of two independent variables where the ``Wronskian'' method fails; 2) there are 
a number of examples of operators of third order of two independent variables which admit invertible Darboux transformations. In this case we cannot pose the question whether
these invertible transformations are a product of Laplace transformations or not, since Laplace transformations only defined for the equation mentioned above.

In~\cite{shem:darboux2} we suggested an algebraic formalism for Darboux transformations. 
In~\cite{shem_2015_type_I} for operators of arbitrary orders and depending in arbitrary number of independent variables we singled out a class of invertible Darboux transformations -- the \textit{Darboux transformations of type I}.
The inverse of Darboux transformations of type I can be described by compact formulas on the operators level. Following an analogy with Laplace transformations,
in this paper we describe the orbits of Darboux transformations of type I in terms of differential invariants.

We consider Darboux transformations of type I for hyperbolic operators $L$ of order three in two independent variables. 
We obtain the following results: \\
1) If we fix the principal symbol of auxiliary operator $M$ (for example, $\sigma(M)=p_x$)
and consider all possible Darboux transformations of type I with $M=\partial_x+m$, where $m$ is a parameter, then 
all of them just transform $L$ into operators that differ by a gauge transformations. We obtained the exact formulas in terms of differential invariants using 
the generating set found in~\cite{invariants,invariants_gen}. \\
2) We observed that Darboux transformations of type I have their own invariants that can be given in terms of gauge invariants of $L$. \\
3) We discovered the first example of a family of operators of order three for which there is an analogue of the Laplace infinite chain, and which we can solve using our method,
while the major Computer Algebra Systems cannot. \\
4) We obtained a number of examples of orbit with non-trivial topological structure. 

Overall, this work announces the first interesting progress on the orbits of Darboux groupoid. 

\section{Groupoid of Darboux transformations} 
Our approach is algebraical: we let $K$ to be a differential field of characteristic zero with commuting derivations $\partial_x, \partial_y$, and $K[D]=K[D_x, D_y]$ to be the corresponding ring of linear
partial differential operators over $K$, where $D_x, D_y$ correspond to derivations $\partial_x, \partial_y$. 
\begin{definition} Consider category $Dar(S)$: \\
(1) The objects of $Dar(S)$ are operators from $\Kall$ with the same arbitrary but fixed principal symbol $S$. \\
(2) Pair $({M},{N})$ is a morphism with source ${L}$ and target ${L}_1$ if 
\[
{N} {L} = {L}_1 {M} \quad \text{up to equivalence} \quad ({M},{N}) \sim ({M} +{A} {L} \  , 
{N} + {L}_1 {A} ) \, ,
\]
where ${A} \in \Kall$ is arbitrary. Another notation for this morphism indicating the source and target, is $\t{L}{(M,N)}{L_1}$. \\
(3) Consecutive morphisms can be composed:\\
\indent $(M,N)\cdot(M_1,N_1) =(M_1 M ,N_1 N)$. \\
(4) For any object $L$ \textit{the identity morphism} is $1_L= (1,1)$.

The morphisms are called \textit{Darboux transformations}. 
\end{definition}
In~\cite{shem:darboux2} it was proved that the composition of morphisms is well defined.
It follows that if a Darboux transformation $(M,N)$ with source $L$ and target $L_1$ has an inverse $(M',N')$, then for some $A,G \in \Kall$ 
the following equalities hold:
\begin{align}
&{{M'M=1+AL}} \ , \label{inv:prop1} \\
&{{MM'=1+GL_1}} \ , \label{inv:prop2} \\  
&N'N=1+LA \ , \label{inv:nl} \\
&NN'=1+L_1G \ .  \label{inv:nl1}
\end{align}

In~\cite{shem_2015_type_I} we showed that axillary operators $A$ and $G$ are related:
\begin{equation*} \label{eq:gn_is_ma}
 GN =MA \ ,
\end{equation*}
and that two of the four conditions above are redundant: \eqref{inv:nl} and \eqref{inv:nl1} follow from~\eqref{inv:prop1} and \eqref{inv:prop2}.

In general case, the automorphisms in this category are not trivial.
\section{Darboux transformations of type I}
In~\cite{shem_2015_type_I} we singled out a class of Darboux transformations which can be completely described on the level of operators:
it consists of Darboux transformations $(M,N)$ with source $L$ such that
\[
 L=CM+f \, , \  f \in K \, .
\]
Here $L,M \in \Kall$. These transformations are always invertible, and the explicit formulas for the target $L_1$, $N$ and its inverse $(M',N')$ are
 \[
  L_1 =M^{1/f} C+f \, , \quad N = M^{1/f} \, , \ M'  = - \frac{1}{f} C \, , \quad N' =-C \frac{1}{f} \, , 
 \]
and the axillary operators from~\eqref{inv:prop1} and~\eqref{inv:prop2} are $A=G=1/f$.
We shall refer to such $(M,N)$ as to \textit{Darboux transformations of type I}.

This class is large enough. Indeed, Laplace transformations, and every invertible Darboux transformation of first order (where the order of a Darboux transformation $(M,N)$ is the order of $M$ and $N$)
is of type I.  
\section{Darboux transformations for gauge differential invariants}
For operators of the form 
\begin{equation} \label{eq:main:class_xys}
L= \partial_x \partial_y (\partial_x + \partial_y) + a_{ij} \partial_x^i \partial_y^j \ ,
\end{equation}
where we use multi-index notation, $a_{ij} \in K$, $i,j=0,1,2$ under gauge transformations  $\displaystyle L \rightarrow e^{-g} L e^g$, $g \in K \setminus \{0\}$,
a generating set for differential invariants is known~\cite{invariants,invariants_gen}:
\[
\begin{array}{ll}
 I_1 &= -2a_{20}+a_{11}-2a_{02} \, ,\\
 I_2 &= \partial_x(a_{20}) -\partial_y(a_{02}) \, ,\\
 I_3 &= a_{10}+ a_{20}(a_{20}-a_{11}) +\partial_y(a_{20}-a_{11}) \, ,\\
 I_4 &= a_{01}+a_{02} (a_{02}-a_{11}) +\partial_x(a_{02}-a_{11}) \, , \\
 I_5 &= a_{00}-a_{01} a_{20}-a_{10} a_{02} + a_{02} a_{20} a_{11}+  \\
  &+( 2 a_{02} - a_{11} +2 a_{20}) \partial_x(a_{20}) + \partial_{xy}(a_{20}-a_{11}+a_{02}) \, .\\
\end{array}
\]

\begin{lemma} Operator of form~\eqref{eq:main:class_xys} has an invertible Darboux transformation 
\begin{enumerate}
 \item With some $M=\partial_x+m$ if and only if $I_4+2I_2=I_{1x}$;
 \item With some $M=\partial_y+m$ if and only if $I_{1y}+2I_2=I_3$; 
 \item With some $M=\partial_x+\partial_y+m$ if and only if $I_3-I_4=I_2$.
\end{enumerate}
Here  $m \in K$.
\end{lemma}

\begin{lemma} 
\label{lem:invS:dx} 
Invertible Darboux transformations for~\eqref{eq:main:class_xys} with $M$ of the form $M=\partial_x+m$ can be 
lifted to gauge invariants: $(I_1,I_2,I_4,I_4,I_5) \mapsto (J_1,J_2,J_3,J_4,J_5)$, where   
\begin{align*}
 J_1 &=I_1 -T_x \, , \\
 J_2 &=I_2 + T_{xy} \, , \\
 J_3 &=I_3 + I_2 + T_{xy} \, , \\
 J_4 &=0 \, , \\
 J_5 &=f - I_{4y}/2 + I_{3x} - I_{2y} +T_{xxy}/2 \, ,
\end{align*}
where $f=I_5-I_1 I_2 - I_{4y}/2$, $T=\ln(f)$. Here $f \neq 0$ and $L=CM+f$ for some $C \in \Kall$.

With $M$ of the form $M=\partial_y+m$:
\begin{align*}
 J_1 &=I_1 -T_y \, , \\
 J_2 &=I_2 - T_{xy} \, , \\
 J_3 &=0 \, , \\
 J_4 &=I_4 -I_2 +  T_{xy} \, , \\
 J_5 &=I_{1} I_2 - I_1 T_{xy} - I_2 T_y - I_{3x}/2 + I_{2x} +I_{4y} +T_{xyy}/2 +T_{xy} T_y +f \, ,
\end{align*}
where $f=I_5- I_{3x}$, $T=\ln(f)$. Here $f \neq 0$ and $L=CM+f$ for some $C \in \Kall$.

With $M$ of the form $M=\partial_x+\partial_y+m$:
\begin{align*}
 J_1 &=I_1 +T_x+T_y \, , \\
 J_2 &=I_2  \, , \\
 J_3 &=-I_4+2I_3 + I_{1y} + T_{xy} + T_{yy} \, , \\
 J_4 &=I_{1x}+2I_4 -I_3 + T_{xx} + T_{xy} \, , \\
 J_5 &=I_{4y} + I_{3x} + I_{1xy}/2 -I_1 I_4 - T_x I_4 -T_y I_4 +T_{xxy}/2 +T_{xyy}/2 + f\, ,
\end{align*}
where $f=I_5+ I_1 I_4 + I_{1xy}/2$, $T=\ln(f)$. Here $f \neq 0$ and $L=CM+f$ for some $C \in \Kall$.
\end{lemma}
\begin{lemma} \label{lem:dinv} The following are invariants of invertible Darboux transformations:
\begin{align*}
 A_x=I_2+I_{1y} & \ , \quad \text{in case} \quad \sigma(M)=p_x \, , \\
 A_y=I_2-I_{1y} & \ , \quad \text{in case} \quad \sigma(M)=p_y \, , \\
 I_2 & \ , \quad \text{in case} \quad \sigma(M)=p_x+p_y \, .
\end{align*}
\end{lemma}%
%
\section{New integrable PDE(s) of third order}
From conditions above, it follows that if an infinite sequence of Darboux transformations of type I with $\sigma(M)=p_x$ exists,
then starting from at least the second term of the sequence $I_4=0$, and therefore, $2I_2=I_{1x}$. 

For the example, we set $I_4=0$, $2I_2=I_{1x}$, and $I_1=2x-y$, $I_2=I_3=1$, $I_5=0$. In particular, 
$L_0= \partial_x \partial_y (\partial_x + \partial_y) -y \partial_x \partial_y + (x-y) \partial_y^2  -x^2+xy-1$ belongs to this gauge equivalence class.
This operator has two different factorizations into three factors each. The factors have symbols $p_y, p_x, p_x+p_y$ and $p_x, p_x+p_y, p_y$, reading from the left to the right. 
Thus, the partial differential equation corresponding to $L_0$ is easy to solve provided we know how to solve first-order partial differential equations in our setting.

On the other hand, there is an infinite chain of invertible Darboux transformations with $\sigma(M)=p_x$ starting at this operator.
The $n$-th term, $n \geq 0$, of this chain has invariants
\begin{align*}
 I^n_1 &=2x-y-2n/(2x-y) \, , \\ 
 I^n_2 &=1+2n/(2x-y)^2 \, , \\ 
 I^n_3 &= n+1+n(n+1)/(2x-y)^2 \, , \\ 
 I_4^n &=0 \, ,
 \\ I_5^n &=-2nx+ny-4n^2/(2x-y)^3 \, ,
\end{align*}
where $n$ is an index. 

\noindent
1) Any partial differential equation having the corresponding values of five invariants can be solved in quadratures using our method. \\
2) Starting from the second term in the chain none of the corresponding operators are factorizable, so we are getting equations of a new, more complicated type. \\
3) For a specific equations, let us consider, for example, one of the simplest that correspond to the values of invariants
obtained on step two (corresponds to $n=1$ in the formulas above):
\begin{align*}
  u_{xxy}+u_{xyy}  &+ \left(x-y - \frac{2}{2x-y}\right) u_{yy}
 -\left(y + \frac{2}{2x-y}\right) u_{xy} \\
 & +u_x + \left(xy-x^2 + \frac{y}{2x-y}\right) u_y
 -x -\frac{2}{2x-y} = 0 \, .
\end{align*}

The latest version (18) of one of the leading Computer Algebra System MAPLE cannot solve this equation.
\section{Orbits structures examples}

The structure of the orbit of the famous Laplace transformations is just one chain, infinite or finite.
Their analogues for operators of higher order -- Darboux transformations of type I appear to generate much more interesting 
structures. 

Below we show a number of examples. In the vertices of the graphs below are the values of the five gauge invariants.
Over the arrows we indicate the principal symbol of the auxiliary operator $M$. If there is no arrow then there is
no corresponding Darnoux transformation. In other words, all invertible Darboux transformations of order one are shown here.

I. Orbits of finite length:
\begin{center}
\begin{tikzpicture}[node distance=4cm, auto]
  \node (L) {$(0,0,0,0,e^{g(x)})$};
  \node (L1) [right of=L] {$(-g_x,0,0,0,e^{g(x)})$};
  \node (L2) [left of=L] {$(g_x,0,0,g_{xx},e^{g(x)})$};
  \draw[->,bend left] (L) to node {$p_x+p_y$} (L2);
  \draw[->] (L2) to node {$p_x$} (L);
  \draw[->] (L) to node {$p_x$} (L1);
  \draw[->,  bend left] (L1) to node {$p_x+p_y$} (L);
   \path (L) edge [loop above] node {$p_y$} (L);
   \path (L1) edge [loop above] node {$p_y$} (L1);
     \path (L2) edge [loop above] node {$p_y$} (L2);
\end{tikzpicture}
\end{center}
\begin{center}
\begin{tikzpicture}[node distance=4cm, auto]
  \node (L) {$(0,0,0,0,x)$};
  \node (L1) [right of=L] {$(-1/x,0,0,0,x)$};
  \node (L2) [left of=L] {$(1/x,0,0,-1/x^2,x)$};
  \draw[->] (L) to node {$p_x$} (L1);
  \draw[<->] (L) to node {$p_x$} (L2);
  \draw[->,  bend left] (L1) to node {$p_x+p_y$} (L);
   \path (L) edge [loop above] node {$p_y$} (L);
   \path (L1) edge [loop above] node {$p_y$} (L1);
     \path (L2) edge [loop above] node {$p_y$} (L2);
\end{tikzpicture}
\end{center}
\begin{center}
\resizebox{12cm}{!}{
\begin{tikzpicture}[node distance=5cm, auto]
  \node (L) {$(0,0,0,0,xy)$};
  \node (L1) [right of=L] {$(-1/x,0,0,0,xy)$};
  \node (L2) [above of=L1] {$(-1/x-1/y,0,0,0,xy)$};
  \node (L3) [below of=L] {$(1/y,0,-1/y^2,0,xy)$};
  \node (L4) [right of=L3] {$(1/y-1/x,0,-1/y^2,0,xy)$};
  \node (L5) [above of=L] {$(-1/y,0,0,0,xy)$};
  \node (L6) [left of=L3] {$(1/x+1/y,0,-1/y^2,-1/x^2,xy)$};
  \node (L7) [left of=L] {$(1/x,0,0,-1/x^2,xy)$};
  \node (L8) [above of=L7] {$(-1/y+1/x,0,0,-1/x^2,xy)$};
  \draw[->,line width=0.3mm] (L) to node {$p_x$} (L1);
  \draw[->,line width=0.3mm] (L1) to node {$p_y$} (L2);
  \draw[->,line width=0.3mm] (L1) to node {$p_x+p_y$} (L3);
  \draw[->,line width=0.3mm] (L3) to node {$p_x$} (L4);
  \draw[->,line width=0.3mm] (L) to node {$p_y$} (L5);
  \draw[->,line width=0.3mm] (L3) to node {$p_y$} (L);
  \draw[->,line width=0.3mm] (L5) to node {$p_x$} (L2);
  \draw[->,line width=0.3mm] (L4) to node {$p_y$} (L1);
  \draw[->,line width=0.3mm] (L) to node {$p_x+p_y$} (L6);
  \draw[->,line width=0.3mm] (L6) to node {$p_x$} (L3);
  \draw[->,line width=0.3mm] (L6) to node {$p_y$} (L7);
  \draw[->,line width=0.3mm] (L7) to node {$p_x$} (L);
  \draw[->,line width=0.3mm] (L7) to node {$p_y$} (L8);
  \draw[->,line width=0.3mm] (L8) to node {$p_x$} (L5);
  \draw[->,line width=0.3mm] (L5) to node {$p_x+p_y$} (L7);
  \draw[->,line width=0.3mm] (L2) to node [swap] {$p_x+p_y$} (L);
\end{tikzpicture}
}
\end{center}

II. Infinite orbits:
\begin{center}
\resizebox{12.0cm}{!}{
\begin{tikzpicture}[node distance=5cm, auto]
  \node (L) {$(0,0,0,0,e^{x+y})$};
  \node (L1) [right of=L] {$(-1, 0, 0, 0, e^{x+y})$};
  \node (L2) [right of=L1] {$\dots$};
  \node (L_1) [left of=L] {$(1, 0, 0, 0, e^{x+y})$};  
  \node (L_2) [left of=L_1] {$\dots$};
  \draw[->, line width=0.3mm] (L) to node [swap] {$p_x$} (L1);
  \draw[->, line width=0.3mm] (L) to node {$p_y$} (L1);
  \draw[->, line width=0.3mm] (L1) to node [swap] {$p_x$} (L2);
  \draw[->, line width=0.3mm] (L1) to node {$p_y$} (L2);
  \draw[->, line width=0.3mm] (L_2) to node [swap] {$p_x$} (L_1);
  \draw[->, line width=0.3mm] (L_2) to node {$p_y$} (L_1);%
  \draw[->, line width=0.3mm] (L_1) to node [swap] {$p_x$} (L);
  \draw[->, line width=0.3mm] (L_1) to node {$p_y$} (L);%
  \draw[->, bend left, line width=0.3mm] (L2) to node {$p_x+p_y$} (L);
  \draw[->, bend left, line width=0.3mm] (L1) to node {$p_x+p_y$} (L_1);
  \draw[->, bend left, line width=0.3mm] (L) to node {$p_x+p_y$} (L_2);
\end{tikzpicture}
}
\end{center}

\begin{center}
\resizebox{11.0cm}{!}{
\begin{tikzpicture}[node distance=4cm, auto]
  \node (L) {$(0,0,x,0,1)$};
  \node (L1) [right of=L] {$(0,0,x,0,2)$};
  \node (L2) [right of=L1] {$(0,0,x,0,3)$};
  \node (L3) [right of=L2] {$\dots$};
  \draw[->] (L) to node {$p_x$} (L1);
  \draw[->] (L1) to node {$p_x$} (L2);
  \draw[->] (L2) to node {$p_x$} (L3);
\end{tikzpicture}
}
\end{center}
\begin{center}
\resizebox{12cm}{!}{
\begin{tikzpicture}[node distance=5cm, auto]
  \node (L) {$(0,0,0,0,x+y)$};
  \node (L1) [right of=L] {$(-d, -d^2, -d^2, 0, x+y+d^3)$};
  \node (L2) [above of=L1] {$(-2d,0,0,0,x+y+2d^3)$};
  \node (L3) [above of=L] {$\displaystyle \left(-d, 2d^2,0, -2d^2,x+y \right)$};
  \node (L4) [left of=L] {$(d,d^2,0,-3d^2,x+y+4d^3)$};
  \node (L5) [below of=L] {$(d,-d^2,-3d^2,0,x+y+3d^3)$};
  \node (L6) [left of=L5] {$(2d,0,-2d^2,-2d^2,x+y+2d^3)$};
  \node (L8) [below left of=L6] {$(4d,0,-6d^2,-6d^2,x+y+20d^3)$};
  \node (L9) [below left of=L8] {$(6d,0,-12d^2,-12d^2,x+y+66d^3)$};
  \node (L10) [below left of=L9] {$\dots$};
  \draw[->,line width=0.3mm] (L) to node {$p_x$} (L1);
  \draw[->,line width=0.3mm] (L1) to node {$p_y$} (L2);
  \draw[->,line width=0.3mm] (L) to node {$p_y$} (L3);
  \draw[->,line width=0.3mm] (L3) to node {$p_x$} (L2);
  \draw[->,line width=0.3mm] (L2) to node {$p_x+p_y$} (L);
  \draw[->,line width=0.3mm] (L3) to node {$p_x+p_y$} (L4);
  \draw[->,line width=0.3mm] (L4) to node {$p_x$} (L);
  \draw[->,line width=0.3mm] (L1) to node {$p_x+p_y$} (L5);
  \draw[->,line width=0.3mm] (L5) to node {$p_y$} (L);
  \draw[->,line width=0.3mm] (L) to node {$p_x+p_y$} (L6);
  \draw[->,line width=0.3mm] (L6) to node {$p_x$} (L5);
  \draw[->,line width=0.3mm] (L6) to node {$p_y$} (L4);
  \draw[->,line width=0.3mm] (L6) to node {$p_x+p_y$} (L8);
  \draw[->,line width=0.3mm] (L8) to node {$p_x+p_y$} (L9);
  \draw[->,line width=0.3mm] (L9) to node {$p_x+p_y$} (L10);
\end{tikzpicture}
}
\end{center}

\end{document}